\documentclass[twocolumn,twoside,slac_two,floatfix]{revtex4}
\usepackage{graphicx}
\usepackage{fancyhdr}
\begin{document}

\title{The Inner Jet of the Quasar PKS~1510$-$089 as Revealed by Multi-waveband Monitoring}

%

\author{Alan P. Marscher, Svetlana G. Jorstad, Francesca D. D'Arcangelo, Dipesh Bhattarai,
Brian Taylor, Alice R. Olmstead, and Emily Manne-Nicholas}
\affiliation{Institute for Astrophysical Research, Boston University,
725 Commonwealth Avenue, Boston, MA 02215 USA} \email{marscher@bu.edu}
\author{Valeri M. Larionov, Vladimir A. Hagen-Thorn, Tatiana S. Konstantinova, Elena G.
Larionova, Liudmila V. Larionova, Daria A. Melnichuk, Dmitry A. Blinov, Evgenia N.
Kopatskaya, and Ivan S.Troitsky}
\affiliation{Astronomical Institute, St. Petersburg State University, Universitetskij Pr. 28,
Petrodvorets, 198504 St. Petersburg, Russia; and Isaac Newton Institute of Chile, St. Petersburg
Branch, Russia}
\author{Iv\'an Agudo, Jos\'e L. G\'omez, and Mar Roca-Sogorb}
\affiliation{Instituto de Astrof\'{\i}sica de Andaluc\'{\i}a, CSIC, Apartado 3004, 18080,
Granada, Spain}
\author{Paul S. Smith and Gary D. Schmidt}
\affiliation{Steward Observatory, University of Arizona, Tucson, AZ 85721-0065}
\author{Omar Kurtanidze, Maria G. Nikolashvili, Givi N. Kimeridze, and Lorand A. Sigua}
\affiliation{Abastumani Astrophysical Observatory, Mt. Kanobili, Abastumani, Georgia}

\begin{abstract}
As part of our comprehensive long-term multi-waveband monitoring of 34 blazars, we followed the
activity in the jet of the blazar PKS~1510$-$089 during major outbursts during the first half of
2009. The most revealing event was a two-month long outburst that featured a number of $\gamma$-ray
flares. During the outburst, the position angle of optical linear polarization rotated by about
$720^\circ$, which implies that a single emission feature was responsible for all of the flares
during the outburst. At the end of the rotation, a new superluminal knot ($\sim 22c$) passed through
the ``core'' seen on 43 GHz VLBA images at essentially the same time as an extremely sharp,
high-amplitude $\gamma$-ray and optical flare occurred. We associate the entire multi-flare outburst with
this knot. The ratio of $\gamma$-ray to synchrotron integrated flux indicates that some of the
$\gamma$-ray flares resulted from inverse Compton scattering of seed photons outside the ultra-fast
spine of the jet. Because many of the flares occurred over time scales of days or even hours, there
must be a number of sources of IR-optical-UV seed photons --- probably synchrotron
emission --- surrounding the spine, perhaps in a slower sheath of the jet.
\end{abstract}

\maketitle

\thispagestyle{fancy}


\section{Introduction} 
There are four primary methods for probing the structure and physics of relativistic jets in
blazars on parsec and sub-parsec scales: VLBI imaging in both total and polarized intensity
at millimeter wavelengths, variability of the flux at radio through $\gamma$-ray frequencies,
polarization at radio through optical wavebands, and the spectral energy distribution (SED).
In the past, efforts to study this capricious class of objects have been limited by datasets
with substantial gaps in either time or frequency coverage. Starting in the mid-1990s, this
situation has been greatly alleviated by the availability of instruments such as the Very Long
Baseline Array (VLBA), the {\it Rossi} X-ray Timing Explorer (RXTE), and, most recently, the
{\it Fermi} Gamma-ray Space Telescope.  Together with concerted efforts with optical/near-infrared, 
millimeter-wave, and radio telescopes, these facilities now allow long-term, comprehensive
multi-waveband monitoring of a number of blazars. Such programs are providing valuable insights
into the processes by which jets form and propagate, as well as the locations and physics of flux
outbursts in blazars \citep[e.g.,][]{Mar08,Chat08,Lar08}.

We are leading a monitoring program that provides the data
needed to combine the techniques listed above in order to locate the sites of high-energy emission, 
determine the processes by which X-rays and $\gamma$-rays are produced, and infer the physical
conditions of the jet, including the geometry of its magnetic field. Because of limitations in
sensitivity and available observing time, such comprehensive monitoring can be mounted
only for a relatively small number of objects, 34 in the case of our program. Progress toward an
overall understanding of blazars therefore involves generalization of inferences drawn from the
well-studied specimens into a framework that can be applied to the entire class.

We present here a rich set of observations of the quasar PKS~1510$-$089 ($z$ = 0.361) that
allows us to locate the sites of $\gamma$-ray flares, leading to inferences on the
high-energy emission processes. PKS~1510$-$089 possesses all of the
characteristics of blazars: flat radio spectrum, apparent superluminal motion --- among the fastest
\citep[as high as $45c$ for H$_\circ$=71 km s$^{-1}$Mpc$^{-1}$ and the concordance
cosmology][]{Spergel07} of all blazars observed thus far \citep{J05} --- high-amplitude and rapid
flux variability at all wavebands, strong and variable optical linear polarization, and high
$\gamma$-ray apparent luminosity \citep{Hartman99}. The observations of this blazar from our
program allow us to compare motions, linear polarization, and changing flux of features in the
parsec-scale radio jet with flux variability of the entire source at radio, millimeter-wave, IR,
optical, X-ray, and $\gamma$-ray frequencies, and with optical polarization. The relative timing of
the correlated variations thus found probe the structure and physics of the innermost jet regions
where the flow is accelerated and collimated, and where the emitting electrons are energized.

\section{Observational Results}

\begin{figure*}
\includegraphics[scale=0.8, width=155mm]{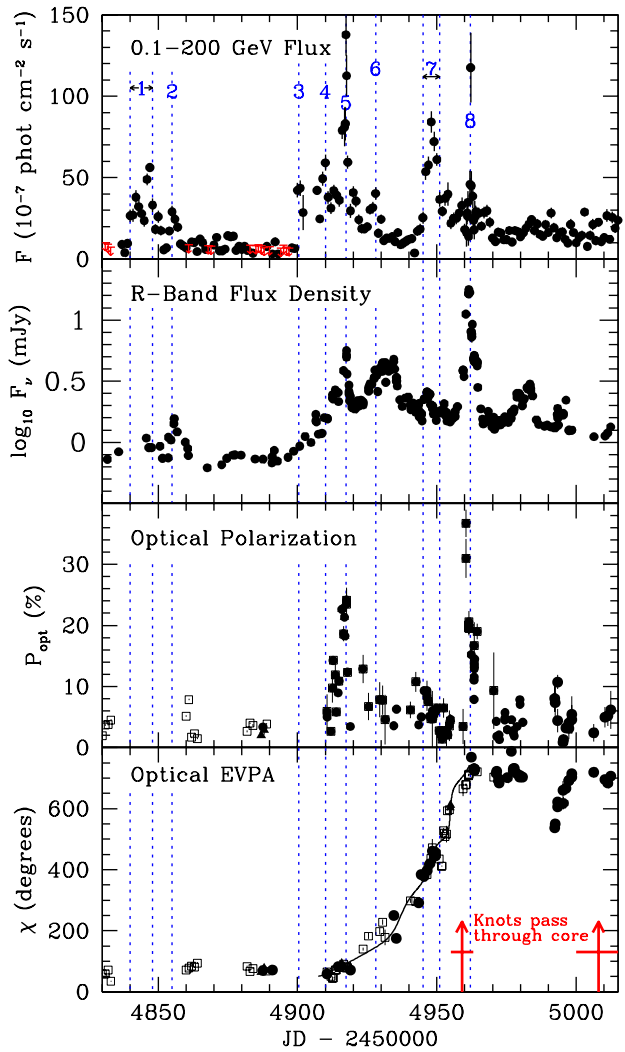}
\caption{Variation with time during the first half of 2009 of (top to bottom panels)
(1) $\gamma$-ray flux (upper limits in red), (2) optical flux density (note the logarithmic
scale, to fit in the very high-amplitude flare on JD 2454962), (3) degree of
optical linear polarization, and (4) electric-vector position angle of optical linear polarization. 
Vertical dotted lines denote major $\gamma$-ray flares. From data presented by \citet{Mar10}.}
\label{fig1}
\end{figure*}

\begin{figure*}
\includegraphics[scale=0.8, width=155mm, clip=true, trim = 0cm 4.2cm 0cm 1.3cm]{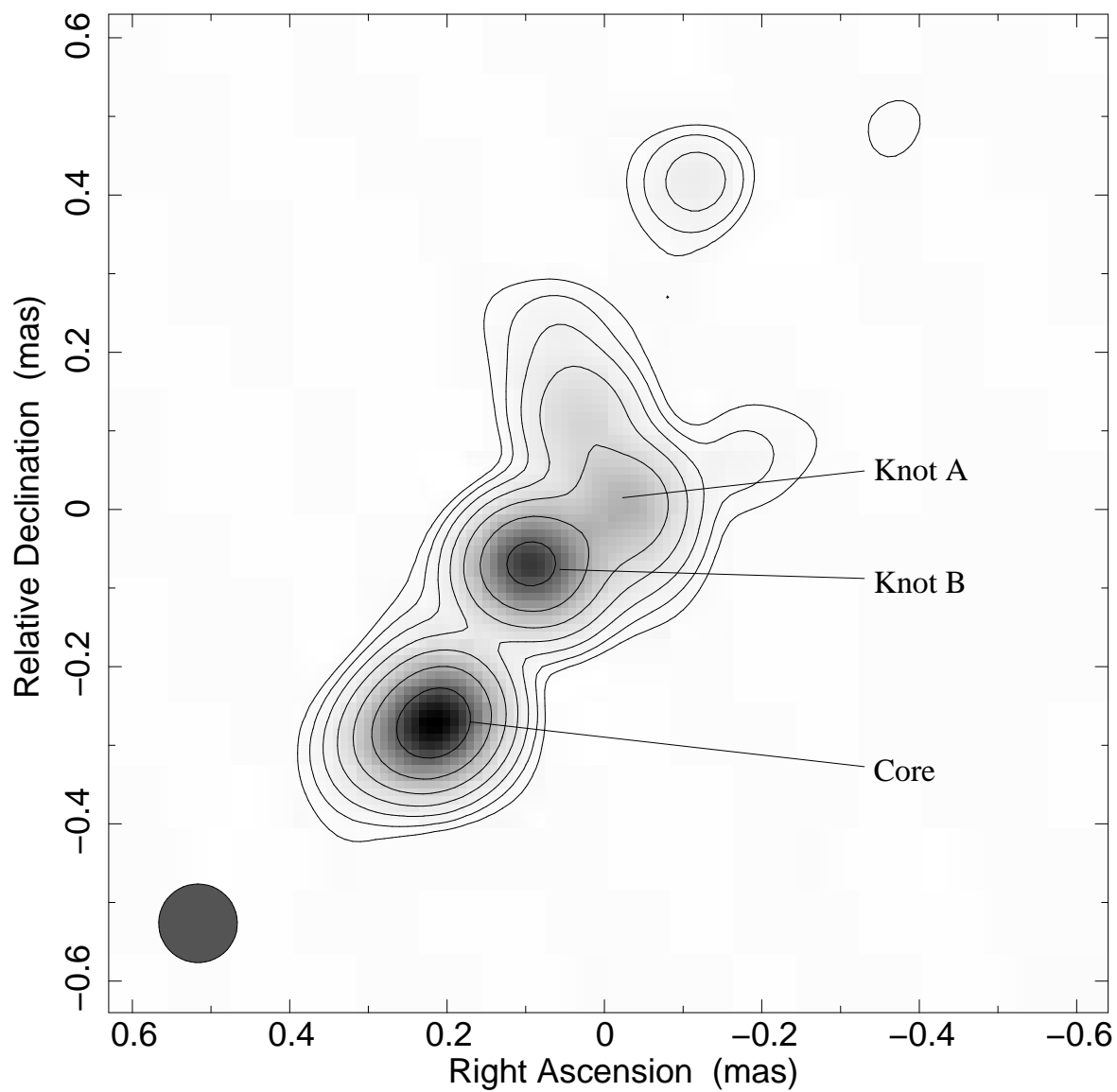}
\caption{VLBA image of PKS~1510$-$089 on 16 September 2009 (JD 2455091). The resolution beam
is comparable to the angular size of the longest baselines of the array in the direction
of the jet. Contour levels are -1, 1, 2, 4, 8, 16, 32, and 64\% of the peak intensity of
0.55 Jy/beam. The two superluminal knots discussed in the text are marked, as is the core.}
\label{fig2}
\end{figure*}

Figure \ref{fig1} presents the {\it Fermi}
Large Area Telescope $\gamma$-ray light curve, the R-band optical light curve, and the
optical polarization as a function of time during the first half of 2009, with the data taken
from the above paper. Details of our observations of PKS~1510$-$089 are given in \citet{Mar10},
which includes data from 2006 to 2009.  That paper also presents 43 GHz VLBA images showing
that a bright superluminal ($22c$) knot (A) passed through the core on JD 2454959$\pm$4.
Additional VLBA images (see Fig. \ref{fig2}) reveal that another knot (B) with the same apparent
speed was coincident with the core on JD 2455008$\pm$8. We stress that the ``core'' of a blazar
seen on VLBI images lies parsecs downstream of the location of the black hole \citep{Mar02}.
There is mounting evidence that, barring optical depth effects, the core represents emission
from jet plasma that has been compressed and energized by a system of one or more
standing conical shocks owing to pressure mismatches between the jet and the external medium.

One of the main questions that our observational program can answer is whether the high-energy
emission occurs between the central engine and core, within the core, or downstream of the core.
In order to do this, we must first establish an association between a superluminal knot and
high-energy flares, then analyze the timing of the flares relative to the motion of the knot.
In the case of BL~Lac in 2005, we employed this method to
determine that there are multiple sites of X-ray and optical
flares in the jet: upstream of the core where the magnetic field has
a helical geometry, and within the core. Flares occur as a disturbance in the jet flow
(an ``emission feature'') passes through these regions \citep{Mar08}.

Eight major flares are apparent in Figure \ref{fig1}, which covers the first half of 2009. The
figure numbers the flares. (We note that $\gamma$-ray emission was detected at an elevated level
throughout the 50+ days following flare 8, but one cannot isolate distinct flares during this
period.) Of particular interest is the 51-day period from JD 2454901 to 2454962 during which
flares 3-8 took place. As seen in Figure \ref{fig1}, the optical polarization position angle
$\chi$ rotated by $\sim~720^\circ$. (Since there is no distinction between $\chi$ and
$\chi\pm 180^\circ n$, we select $n$ such that the jump in $\chi$ is minimized.) The ratio
of $\gamma$-ray 0.1-200 GeV integrated flux to bolometric synchrotron flux \citep[see][]{Mar10}
for flares 1--8 was 70, 30, 40, 40, 30, 10, 40, and 9, respectively.

Superluminal knot A passed through the core at the same time (within the uncertainties) as the
very sharp (intra-day), extremely high-amplitude $\gamma$-ray and optical flare on JD
2454962. The rotation of the optical polarization vector ended at the same time. We therefore
surmise that the $\gamma$-ray flares occurred upstream of and within the core. After this
point, the $\gamma$-ray flux remained elevated, which suggests that $\gamma$-ray emission
continued downstream of the core, as inferred previously from observations with EGRET, the
VLBA, and single-dish radio antennas \citep{J01,LV03}.

\section{Interpretation}

Although a turbulent magnetic field can produce such apparent rotations of the polarization
vector by $720^\circ$ or more via a random
walk \citep{all85,jones88,darc07}, the beginning and end of such a rotation should occur randomly,
whereas the observed event matches quite closely the duration of the flux outburst. We therefore
conclude that a single coherent emission feature was responsible for the entire outburst. We
appeal to the same model as for BL~Lac \citep{Mar08}: as a disturbance in the flow moves down
the jet, it follows a spiral streamline \citep[as predicted for a model in which the jet is
magnetically accelerated and collimated;][]{vlah06}. It first travels through the acceleration
and collimation zone, which possesses a tightly wound helical magnetic field. If the emission
feature covers most, but not all, of the cross-section of the jet, much of the polarization
is cancelled out from mutually perpendicular magnetic field orientations around the
circumference. The residual polarization vector rotates as the feature executes its
spiral motion.

We have performed a rough calculation of the rotation of $\chi$ in such a model, matching the
parameters (probably non-uniquely) to the PKS~1510$-$089 data. The bulk Lorentz
factor increases from 8
to 24 during the rotation event, causing the Doppler factor to increase from 15 to 38 for an
angle between the jet axis and the line of sight of $1.4^\circ$ \citep{J05}. The acceleration
of the jet causes an increase with time in the rate of rotation, as observed. The circulation
of the centroid of the emission feature around the jet axis causes $\chi$ (corrected for
relativistic aberration) to undulate with time about the trend of monotonic rotation. In the
quantitative model, the
core lies 17 pc downstream of the point where the outburst containing flares 3-8 began.
The magnetic field is in the range of 1 to 0.2 G during the outburst.

The dichotomy in the ratio of $\gamma$-ray to synchrotron flux of the various flares suggests
that different sources of seed photons dominate the inverse Compton scattering as the
emission feature moves down the jet. Flares with relatively low values of this ratio can
be caused by an increase in the number of GeV electrons, which causes enhanced synchrotron,
synchrotron self-Compton (SSC), and external Compton (EC) emission. If SSC is the main
mechanism, second-order scattering is probably important, since the lowest $\gamma$-ray to
synchrotron flux ratio, 9, still exceeds unity. There are some flares (1, 3, and 7) where there
is little or no increase in the optical flux. We infer that these correspond to the emission
feature encountering local sources of enhanced IR-optical-UV seed photon density. We
estimate the luminosity of these sources to be $\sim 3\times 10^{43}$ erg/s, too high
to correspond to any likely cosmic source except a slower (but still probably relativistic)
sheath of the jet \citep[see][]{gtc05}. The source could, for example, be a shock in the
outer periphery of the jet.

\begin{figure*}
\includegraphics[width=155mm]{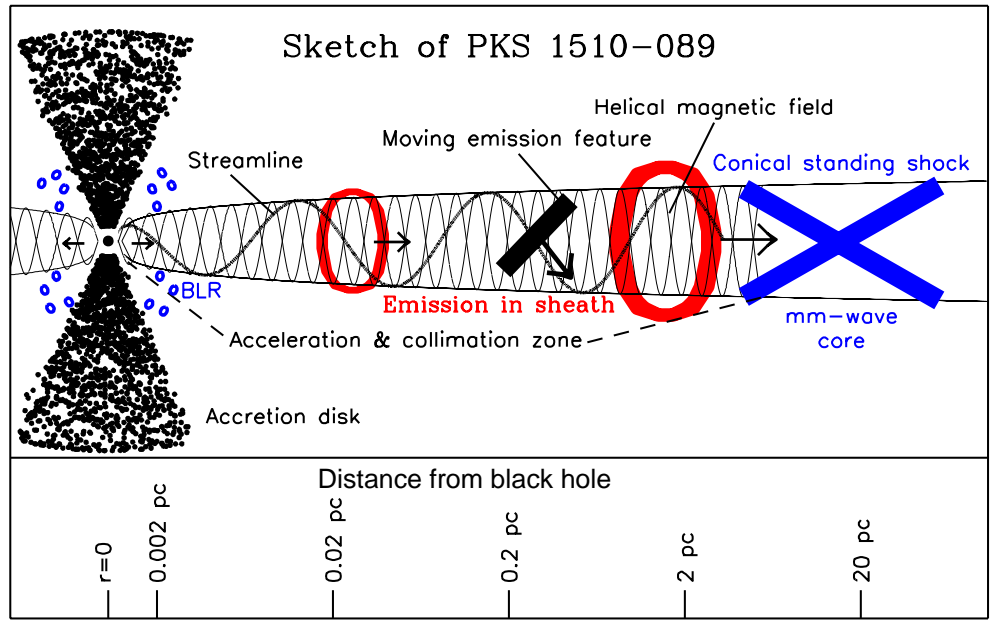}
\caption{Sketch of the model for PKS~1510$-$089 discussed in the text. Note the logarithmic
length scale, so that various components of the nucleus can be included.}
\label{fig3}
\end{figure*}

Figure \ref{fig3} sketches the general picture that we propose. The $\gamma$-ray emission undergoes
flares at various sites along the jet, with events 3-7 representing locations where either
the sheath produces large numbers of seed photons or electrons are accelerated to
GeV energies. Flare 8 takes place as the moving emission feature
becomes compressed and energized by standing shocks in the core. The elevated post-outburst
$\gamma$-ray emission first comes from knot A as it propagates downstream of the core. At
some point --- probably near JD 2454990 --- emission from knot B becomes dominant. This is
weaker than knot A, so the flares are of lower amplitude. But we do see a $\sim~180^\circ$
rotation of the polarization vector {\it in the same direction and with the same slope as
the previous, longer rotation.} This repetition of the pattern is a firm requirement of
our model: the physical structure of the spine or sheath of the jet
could change on a time scale of years, but not months. In contrast, a turbulent field
model predicts that both the sense and rate of rotation should vary from one event to another.

It is possible that flares 1-2 correspond to the disturbance passing through the rich
seed photon field near the accretion disk and within the broad emission-line region (BELR). If
this is the case, then there is an extended region in the jet beyond the BELR where
electrons are not accelerated efficiently to energies exceeding $\sim 1$ GeV.

\section{Conclusions}

We are finally at the point where the richness of our datasets is sufficient for us to draw
grand inferences about the locations and physical mechanisms of $\gamma$-ray flares. If
PKS~1510$-$089 and BL~Lac are typical, it is no wonder that previous, less comprehensive
monitoring programs left us confused! There are multiple sources of seed photons, some of
which are quite local (as opposed to more global, such as an IR-emitting hot dust torus)
and may not radiate at sufficiently high luminosities to be directly observable. These
include the accretion disk and BELR --- which may be important during the earliest part of a
multi-flare outburst --- shocks or other features in the sheath of the jet, and synchrotron
radiation from the fast spine of the jet that contains the electrons that scatter the seed
photons to $\gamma$-ray energies. A single superluminal knot can be responsible for
a number of flares and periods of sustained elevated $\gamma$-ray emission.

As we accumulate data for more blazars, we expect to see similar patterns of flares both
before and after a disturbance passes through the core of the jet. We also hope to
find deviations from this pattern that serve to provide further insight into the range
of physical behavior of blazars. In both cases, we anticipate a tremendous increase
in our understanding of the relativistic jets of blazars by combining the great power
of monitoring the flux, polarization, and sub-milliarcsecond structure of these exciting
objects.
\vspace{-5mm}
\begin{acknowledgments}
This research was funded in part by NASA grants NNX08AV65G, NNX08AV61G,
NNX08AW56G, and NNX08AJ64G, NSF grant AST-0907893,
Russian RFBR grant~09-02-0092, Spanish ``Ministerio de Ciencia e Innovaci\'on'' grant
AYA2007-67626-C03-03, and Georgian NSF grant GNSF/ST08/4-404. The VLBA is an
instrument of the National Radio Astronomy Observatory, a facility of the NSF,
operated under cooperative agreement by Associated Universities, Inc. 
\end{acknowledgments}

\end{document}